\documentclass[a4paper,12pt]{article}

\usepackage[numbers,sort&compress]{natbib}
\usepackage{amsmath,amssymb}
\usepackage{graphicx}
\usepackage{array}
\usepackage{siunitx}
\usepackage{hyperref}

\usepackage{iftex}
\ifXeTeX
  \usepackage{xeCJK}
  \newcommand{\zh}[1]{#1}
\else
  \usepackage{CJKutf8}
  \newcommand{\zh}[1]{\begin{CJK*}{UTF8}{gbsn}#1\end{CJK*}}
\fi

\title{Periodic Orbits of Van der Waals Black Holes%
\thanks{Supported by National Natural Science Foundation of China (Grant No.~12365008),
the Guizhou Provincial Basic Research Program (Natural Science) (Grant No.~QianKeHeJiChu[2024]Young166),
the Guizhou Provincial Basic Research Program (Natural Science) (Grant No.~QianKeHeJiChu-ZK[2024]YiBan027 and
QianKeHeJiChuMS[2025]680), the Guizhou Provincial Major Scientific and Technological Program XKBF (2025) 010
(Hosted by Professor Xu Ning), the Guizhou Provincial Major Science and Technological Program XKGF (2025) 009
(Hosted by Professor Xiang Guoyong) and the Guizhou Provincial Major Scientific and Technological Program
(Hosted by Professor Fan Lulu)}}

\author{Xuyao Gong~(\zh{龚绪尧}), Zhaoyi Xu~(\zh{徐兆意}), Meirong Tang$^{*}$~(\zh{唐美荣})\\[1mm]
College of Physics, Guizhou University, Guiyang 550025, China\\[1mm]
E-mail: tangmr@gzu.edu.cn}

\date{}

\begin{document}
\maketitle

\begin{abstract}
Black holes in asymptotically anti-de Sitter (AdS) spacetime whose
extended-phase-space thermodynamics reproduces that of a van der Waals fluid
exactly form a distinctive class, the van der Waals black holes (VBHs). The
timelike geodesics of VBHs are solved, their periodic orbits are catalogued, and
the gravitational waves emitted along them are computed, so as to trace how the
thermodynamic pressure $P$ and the molecular-volume parameter $b$ shape the
orbital dynamics. Integrating the geodesic equations numerically yields the
energy, angular momentum and radius of the innermost stable circular orbit (ISCO)
and maps out the family of bound orbits. To delimit the parameter window in which
the effective potential stays well behaved, we introduce a critical angular
momentum $L_s$. Within the orbit-classification scheme, we compute the energies
corresponding to rational $q$ for several parameter sets and display the
resulting orbits. The waveforms produced along these orbits are generated within
the ``Kludge'' formalism, which makes it possible to quantify the imprint of $b$
on the wave amplitude and on the period. Both the orbital architecture and the
gravitational-wave signature turn out to be strongly sensitive to $b$. The
dynamical behavior of VBHs and Schwarzschild-AdS (SAdS) black holes differs
markedly, providing a new perspective for probing black hole thermodynamics via
gravitational wave observations.


\end{abstract}

\section{Introduction}
\label{sec:intro}

General relativity established itself early on by resolving, with high precision,
several long-standing discrepancies of Newtonian theory --- the anomalous advance
of Mercury's perihelion and the bending of light near the Sun being the most
celebrated examples \cite{Will2014,GAO2020168194}. The first exact vacuum
solution of Einstein's field equations, the Schwarzschild metric, contains a
black hole, an object that has since attracted sustained attention.
Gravitational-wave observations (e.g., refs.~\cite{LIGOScientific:2016emj,LIGOScientific:2016aoc,LIGOScientific:2020aai,LIGOScientific:2018mvr})
have by now confirmed that black holes are genuine astrophysical objects
\cite{Tu:2023xab}.

Einstein introduced the cosmological constant $\Lambda$ soon after general
relativity, later withdrawing it as superfluous. Since then, $\Lambda$ has
oscillated in and out of favour in the theoretical literature. To this day, it is
widely accepted that the universe possesses a small, positive cosmological
constant $\Lambda$. Yet several theoretical frameworks instead favour
$\Lambda<0$, whose natural vacuum is an asymptotically anti-de Sitter spacetime
\cite{Liang:2023ahd}. Such geometries attract special interest because their
black holes often exhibit phase transitions, the canonical example being the
first-order Hawking-Page transition of Schwarzschild-AdS black holes
\cite{Hawking:1982dh}. Later work in extended-phase-space thermodynamics, which
promotes $\Lambda$ to a thermodynamic pressure $P$
\cite{Altamirano:2014tva,Dolan:2011xt,Dolan:2012jh,Cvetic:2010jb,Larranaga:2011wd,Gibbons:2012ac,Gunasekaran:2012dq,Lu:2012xu,spallucci2013maxwellsequalarealaw,Chen:2013ce,belhaj2014criticalbehaviors3dblack,Altamirano:2013uqa,Belhaj:2013cva,Mo:2013sxa,Mo:2013ela,Ma:2013aqa,Castro:2013pqa,Lu:2013ura,Mo:2014qsa},
showed that charged or rotating AdS black holes behave qualitatively like van der
Waals fluids \cite{Chamblin:1999tk,Cvetic:1999ne,Caldarelli:1999xj,Niu:2011tb,Chamblin:1999hg,Tsai:2011gv,Johnson:2013dka};
for a review see \cite{Kubiznak2017}. Refs.~\cite{Gunasekaran:2012dq,Kubiznak:2012wp}
pushed the analogy further: in the $P$-$V$ plane these black holes display
critical behaviour that mirrors the liquid-gas transition of a van der Waals
fluid, down to the values of the critical exponents
\cite{Gunasekaran:2012dq,Kubiznak:2012wp}. The analogy, however, is only
approximate: the black-hole equation of state departs in detail from the van der
Waals form, and no exact dictionary maps the black-hole parameters (charge $Q$,
angular momentum $J$) onto the fluid ones (attraction $a$, molecular volume
$b$). This limitation was circumvented in ref.~\cite{Rajagopal:2014ewa}, where an
inverse construction was used to build an asymptotically AdS solution whose
extended-phase-space thermodynamics coincides exactly with that of a van der
Waals fluid; the resulting geometry is known as the van der Waals black hole
(VBH) \cite{Rajagopal:2014ewa,Delsate:2014zma}. This model offers a clean setting
for studying black-hole phase transitions in asymptotically AdS spacetimes. Most
work on the VBH so far remains theoretical
\cite{Delsate:2014zma,Pradhan:2016feg,Hu:2017xvm,Ditta:2023luy,Oubagha:2023ghx},
and observational tests are still lacking. A few studies \cite{Molla:2022izk}
probe the model through black-hole shadows, relying mainly on conventional
optical observations. Gravitational-wave observations have not yet been fully
exploited to test the VBH model.

In an extreme-mass-ratio (EMR) binary the companion is so much lighter than the
central supermassive black hole that it can be modelled as a test particle: as it
orbits, it radiates gravitational waves, loses energy and angular momentum, and
gradually spirals inward \cite{Zhao:2024exh,Li:2024tld}. The inspiral sweeps
through a sequence of periodic orbits, each acting as an intermediate stage of
the motion \cite{Zhao:2024exh,Meng2025}. A faithful description of the
gravitational radiation from such systems therefore hinges on a thorough
understanding of these orbits \cite{Glampedakis:2002ya}. Earlier work
\cite{Levin:2008mq,Grossman:2011im,Li:2024tld,Shabbir:2025kqh} has shown that the
properties of these periodic orbits can be exploited to construct gravitational
waveforms for EMR systems in the adiabatic regime \cite{Zhao:2024exh,LIN2021100745}.

This work therefore studies how the thermodynamic parameters --- the pressure $P$
and the van der Waals parameter $b$ --- affect the dynamics of test particles (in
particular their periodic orbits) and the resulting gravitational-wave signals.
The goal is to open a dynamical window onto black-hole thermodynamics and to fill
the gap in observational tests of the VBH model through gravitational waves. The
approach has already proved feasible in many spacetimes
\cite{Tu:2023xab,Li:2024tld,Zi:2023qfk,Zhao:2024exh,Shabbir:2025kqh,Lim:2024mkb,Misra:2010pu,Levin:2009sk,Grossman:2011ps,Lin:2023rmo,Deng:2020yfm,Zhou:2020zys,Wei:2019zdf,Mummery:2022ana,Babar:2017gsg}.

The paper is organised as follows. Section~\ref{sec:geodesics} reviews the VBH
metric and energy conditions and derives the timelike geodesic equations;
section~\ref{sec:bound} examines the bound orbits and their dependence on $P$ and
$b$; section~\ref{sec:periodic} applies the orbit-classification method
\cite{Levin:2008mq}; section~\ref{sec:gw} employs the ``Kludge'' waveform method
\cite{Babak:2006uv} to compute and display the gravitational waveforms radiated
along the periodic orbits and analyses the role of $b$. We conclude in
section~\ref{sec:conclusion}. Except for the waveform plots of
section~\ref{sec:gw}, which use SI units, natural units with $G=c=1$ are adopted
throughout.

\section{VBH and Timelike Geodesics}
\label{sec:geodesics}

Following the inverse procedure of ref.~\cite{Rajagopal:2014ewa} (see also
ref.~\cite{Oubagha:2023ghx}), the VBH is constructed by demanding the required
van der Waals thermodynamics and solving backwards for the asymptotically AdS
metric that produces it,
\begin{equation}
	ds^{2} = - f(r)dt^{2} + \frac{1}{f(r)}dr^{2} + r^{2}d\theta^{2} + r^{2}\sin^{2}\theta d\phi^{2}, \label{eq:metric}
\end{equation}
in the notation of ref.~\cite{Rajagopal:2014ewa}, where $f(r)$ is given by:
\begin{equation}
	f(r) = 2\pi a - \frac{2M}{r} + \frac{r^{2}}{l^{2}}\left( 1 + \frac{3}{2}\frac{b}{r} \right)
	- \frac{3\pi ab^{2}}{r(2r + 3b)} - \frac{4\pi ab}{r}\log{\left( \frac{r}{b} + \frac{3}{2} \right)}. \label{eq:f_original}
\end{equation}
Substituting the parameters characterizing the intermolecular attraction
$a = \frac{1}{2\pi}$, and the thermodynamic pressure
$P = - \frac{\Lambda}{8\pi} = \frac{3}{8\pi l^{2}}$ into the above expression
simplifies $f(r)$ to:
\begin{equation}
	f(r) = 1 - \frac{2M}{r} + \frac{8\pi P}{3}r^{2}\left( 1 + \frac{3}{2}\frac{b}{r} \right)
	- \frac{3}{2}\frac{b^{2}}{r(2r + 3b)} - \frac{2b}{r}\log\left( \frac{r}{b} + \frac{3}{2} \right). \label{eq:f_simplified}
\end{equation}
Throughout, $P$ denotes the thermodynamic pressure and $b$ the molecular-volume
scale \cite{Rajagopal:2014ewa,Oubagha:2023ghx}.

As the spacetime is asymptotically AdS, $P>0$; setting $b=0$ recovers the
Schwarzschild solution, and the positivity of the molecular volume requires
$b\geq 0$. The matter content must obey the energy conditions (see, e.g.,
refs.~\cite{Rajagopal:2014ewa,Oubagha:2023ghx}):
\begin{align}
	&\text{Weak:}\quad \rho \geq 0,\ \ \rho + p_{i} \geq 0,\\
	&\text{Strong:}\quad \rho + \sum_{i}p_{i} \geq 0,\ \ \rho + p_{i} \geq 0,\\
	&\text{Dominant:}\quad \rho \geq |p_{i}|, \label{eq:energy_conditions}
\end{align}
where the various components, taken from ref.~\cite{Rajagopal:2014ewa} (see also
ref.~\cite{Oubagha:2023ghx}), read:
\begin{align}
	&\rho = - p_{1} = \frac{1 - f(r) - rf'(r)}{8\pi r^{2}} + P, \\
	&p_{2} = p_{3} = \frac{rf''(r) + 2f'(r)}{16\pi r} - P. \label{eq:energy_components}
\end{align}
However, due to the logarithmic term in $f(r)$, analytical solution is extremely
difficult, so it is reasonable to find numerical solutions for selected
parameters and discuss them.

\begin{figure}[tbp]
	\centering
	\includegraphics[width=0.8\textwidth]{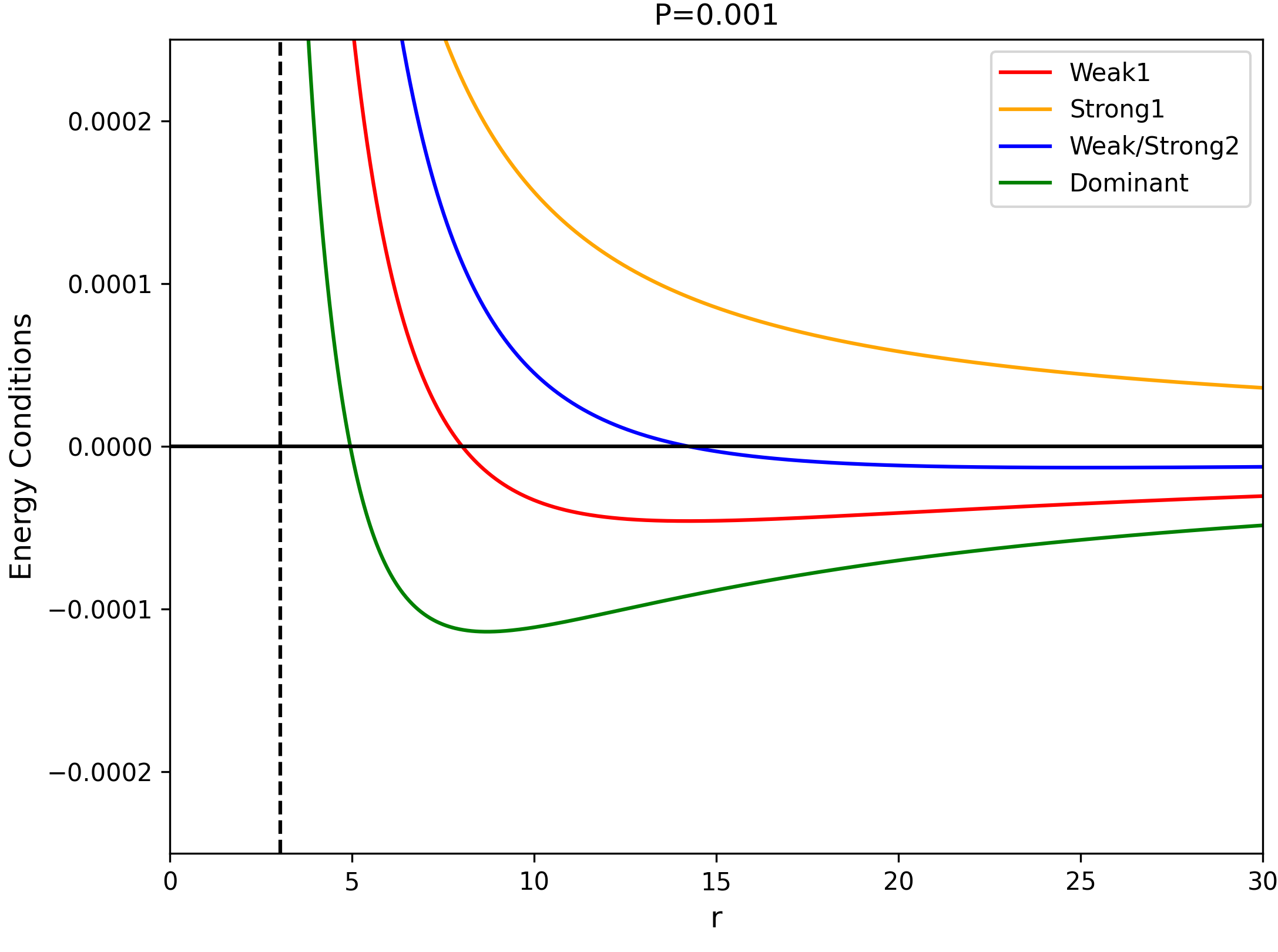}
	\caption{Variation of energy conditions with $r$ for $P = 0.001$. An energy condition is satisfied when both function values are greater than zero (for the dominant energy condition, only one function value need be greater than zero). The dashed line indicates the event horizon radius.}
	\label{fig:energy_conditions_1}
\end{figure}

For $P = 0.001$ (with $b$ and $M$ normalized), figure~\ref{fig:energy_conditions_1}
shows that, moving outward in $r$, the dominant condition fails first, followed
in turn by the weak and the strong conditions.

\begin{figure}[tbp]
	\centering
	\includegraphics[width=0.8\textwidth]{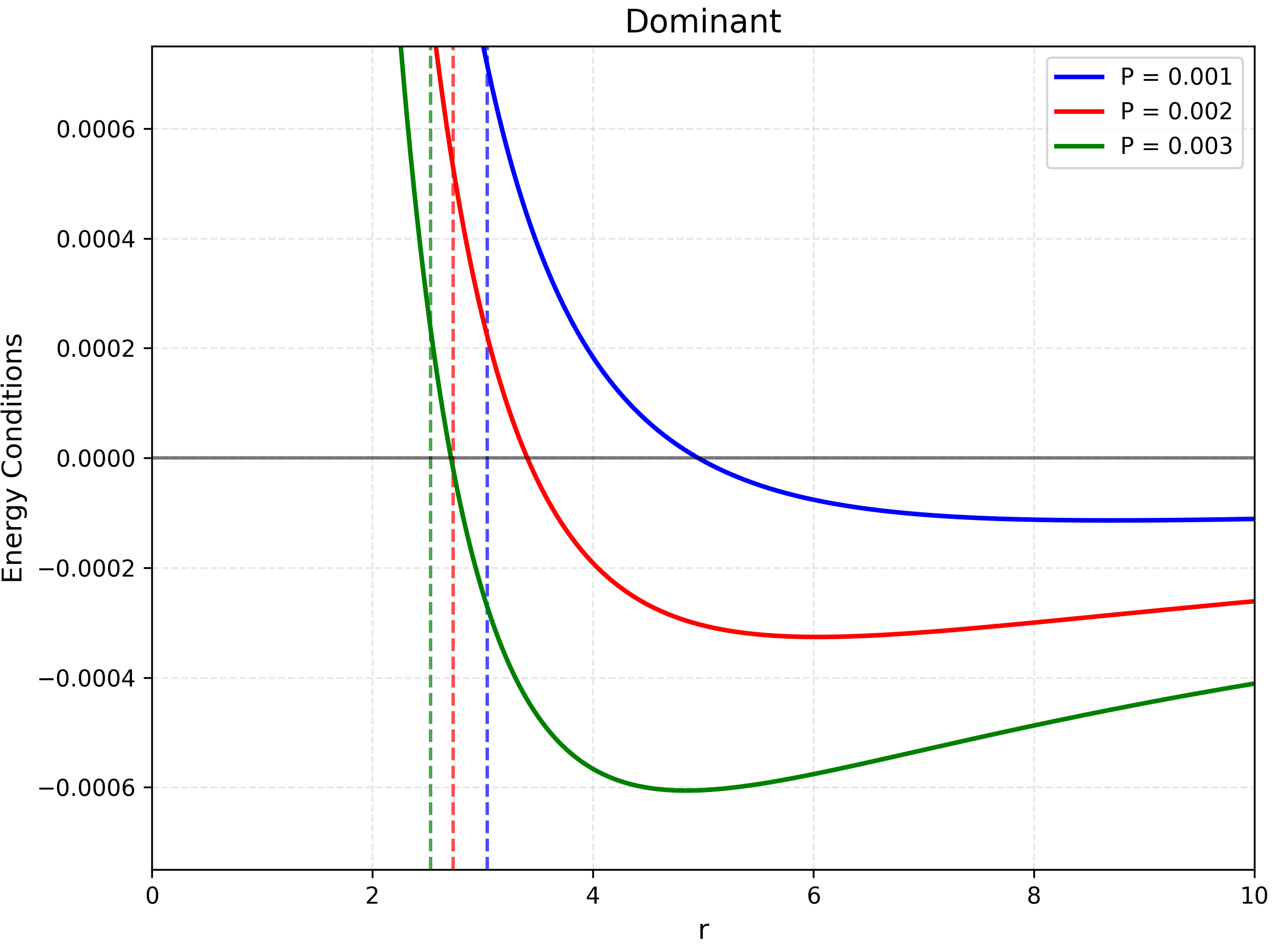}
	\caption{Variation of the dominant energy condition with $r$ for different $P$. The condition is satisfied when the function value is greater than zero. The dashed line indicates the event horizon radius for each $P$.}
	\label{fig:dominant_condition}
\end{figure}

Figure~\ref{fig:dominant_condition} shows that as $P$ grows, the point at which
the dominant energy condition starts to fail moves toward the horizon. Adopting
the approach of ref.~\cite{Hu:2017xvm}, each condition is imposed jointly with
the horizon equation $f(r_+)=0$ at fixed $M>0$, which fixes how $b$ and $P$ scale
with the horizon radius $r_+$ (table~\ref{tab:energy_ratios}). When the ratios
$b/r_{+}$ and $P/r_{+}$ are respectively less than $n_{1}$ and $n_{2}$, a
corresponding number of energy conditions are satisfied (e.g., 1, 2, or all
conditions).

\begin{table}[tbp]
	\centering
	\caption{Relations between parameters and horizon radius for satisfying different numbers of energy conditions.}
	\begin{tabular}{|l|c|c|}
		\hline
		Energy condition & $n_{1}$ ($b/r_{+} < n_{1}$) & $n_{2}$ ($P/r_{+} < n_{2}$)\\
		\hline
		Strong / satisfies 1 condition & 1.11367769 & 0.05794531 \\
		Weak / satisfies 2 conditions & 0.63738977 & 0.03571093 \\
		Dominant / satisfies all conditions & 0.46201214 & 0.02076857 \\
		\hline
	\end{tabular}
	\label{tab:energy_ratios}
\end{table}

Unless noted otherwise, the numerical results below use $P \in \{0.0001, 0.0005,
0.001, 0.002\}$ and $b \in \{0, 0.3, 0.5, 0.7, 1.0\}$, each pair ($P$, $b$) remaining
within the bounds imposed by the energy conditions of table~\ref{tab:energy_ratios}.

In an EMR system the light companion --- say a stellar-mass object --- orbits the
supermassive central body (here a VBH) along a timelike geodesic. The
corresponding Lagrangian reads
\begin{equation}
	\mathcal{L} = -\frac{1}{2}g_{\mu\nu}{\dot{x}}^{\mu}{\dot{x}}^{\nu}, \label{eq:lagrangian}
\end{equation}
where $\dot{x} = dx^{\mu}/d\tau$. Substituting the metric (\ref{eq:metric}) and
choosing the plane $\theta = \pi/2$:
\begin{equation}
	\mathcal{L} = \frac{1}{2}\left[ f(r){\dot{t}}^{2} - \frac{1}{f(r)}{\dot{r}}^{2} - r^{2}{\dot{\phi}}^{2} \right]. \label{eq:lagrangian_2d}
\end{equation}
The symmetries of this static, spherically symmetric metric supply two conserved
quantities, the energy
\begin{equation}
	E = \frac{\partial\mathcal{L}}{\partial\dot{t}} = f(r)\dot{t}, \label{eq:energy}
\end{equation}
and the angular momentum
\begin{equation}
	L = - \frac{\partial\mathcal{L}}{\partial\dot{\phi}} = r^{2}\dot{\phi}; \label{eq:angular_momentum}
\end{equation}
for a timelike geodesic the four-velocity obeys the normalisation
\begin{equation}
	g_{\mu\nu}{\dot{x}}^{\mu}{\dot{x}}^{\nu} = - 1. \label{eq:normalization}
\end{equation}
(see, e.g., ref.~\cite{Liang:2023ahd})

Substituting the metric (\ref{eq:metric}), energy (\ref{eq:energy}), and angular
momentum (\ref{eq:angular_momentum}) yields:
\begin{equation}
	{\dot{r}}^{2} = E^{2} - V_{eff}, \label{eq:radial_equation}
\end{equation}
where $V_{eff}$ is the effective potential:
\begin{equation}
	V_{eff} = f(r)\left( 1 + \frac{L^{2}}{r^{2}} \right). \label{eq:effective_potential}
\end{equation}
For fixed angular momentum $L$, as $r \rightarrow \infty$,
$(1 + L^{2}/r^{2}) \rightarrow 1$, so the behavior of $V_{eff}$ is primarily
determined by $f(r)$. Therefore, the trend of $V_{eff}$ can be inferred from that
of $f(r)$.

\begin{figure}[tbp]
	\centering
	\includegraphics[width=0.8\textwidth]{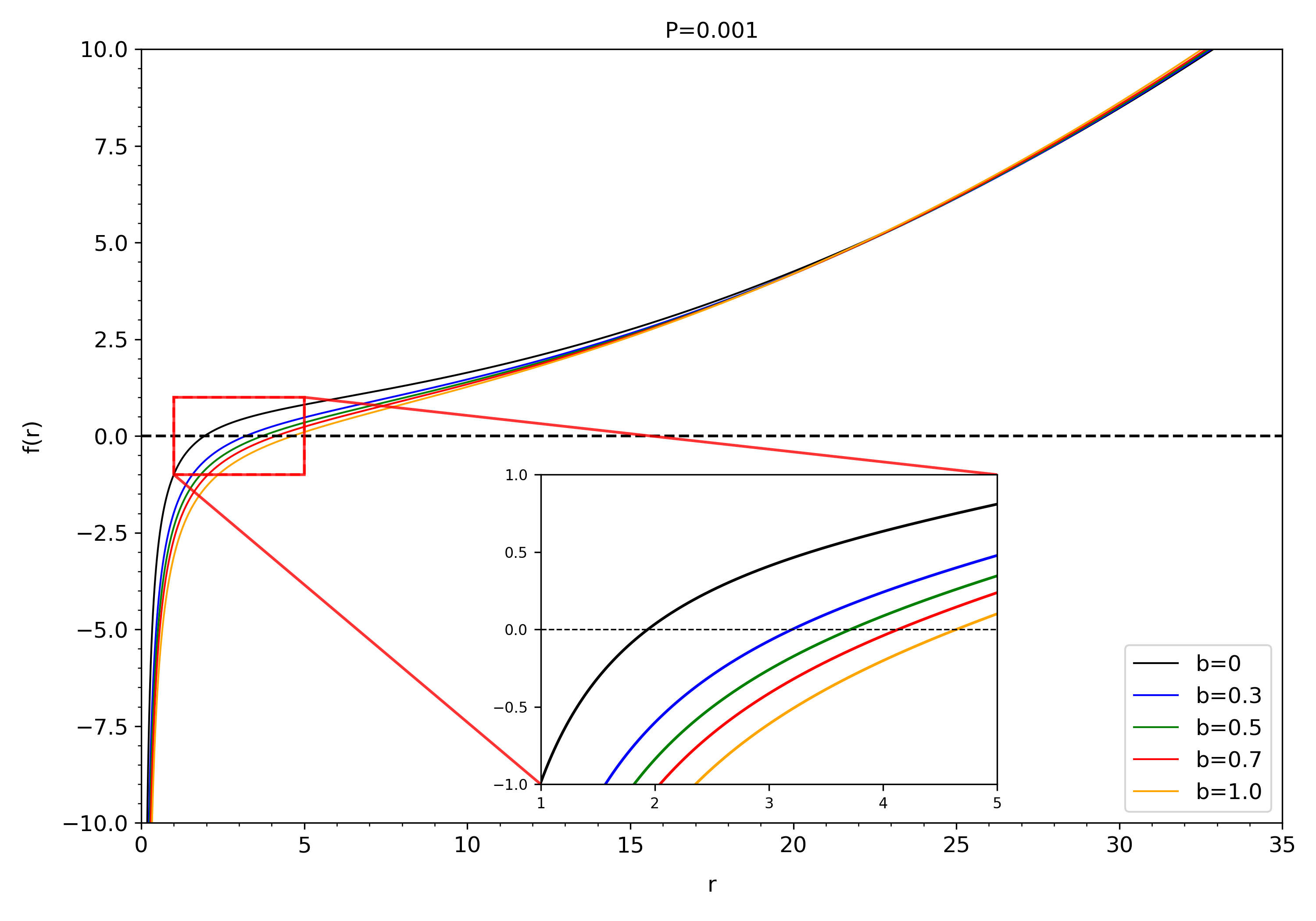}
	\caption{Variation of $f(r)$ with $r$. $f(r)$ tends to infinity as $r$ increases. The event horizon radius $r_{+}$ increases with $b$. For $b=0$, the metric reduces to the SAdS case (with $P=0.001$).}
	\label{fig:f_r}
\end{figure}

Figure~\ref{fig:f_r} shows that in AdS spacetime $f(r)$ grows without bound; the
same holds for $V_{eff}$. The particle is therefore always confined by the VBH
and can never reach infinity: its motion reduces to either a bound orbit or an
infall. In particular, because $V_{eff}$ diverges, a particle with $E>1$ can
still remain bound, since $\dot{r}^{2}$ becomes negative at sufficiently large
$r$ for finite energy. Orbits whose apocenter lies farther out require larger
energies. This hints that asymptotically AdS regions ($\Lambda<0$), even where
they exist, may only be local features of spacetime.

\section{Bound Orbits in VBH Spacetime}
\label{sec:bound}

The analysis starts from the innermost stable circular orbit (ISCO), i.e. the
circular orbit of smallest radius that is still stable. No stable circular motion
exists inside it: there, an arbitrarily small perturbation makes the particle
fall into the black hole \cite{GAO2020168194}.

The conditions for the ISCO are:
\begin{equation}
	V_{eff} = E^{2}, \quad \frac{dV_{eff}}{dr} = 0, \quad \frac{d^{2}V_{eff}}{dr^{2}} = 0. \label{eq:ISCO_conditions}
\end{equation}
Because the logarithmic term in $f(r)$ precludes a closed-form solution,
$r_{ISCO}$, $L_{ISCO}$ and $E_{ISCO}$ are determined numerically.

\begin{figure}[tbp]
	\centering
	\includegraphics[width=0.8\textwidth]{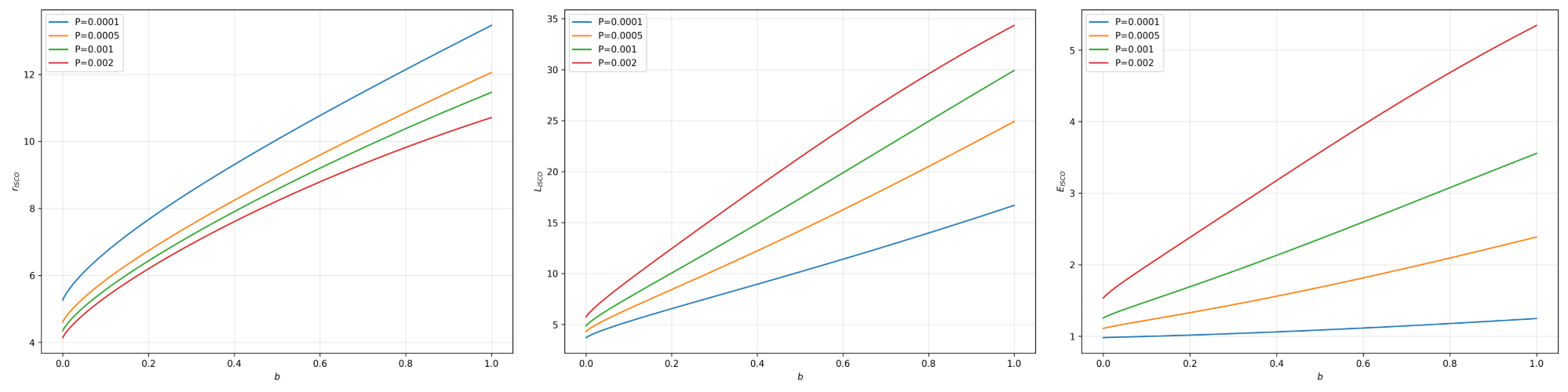}
	\caption{ISCO parameters $r_{ISCO}$, $L_{ISCO}$, and $E_{ISCO}$ as functions of $b$ for various $P$.}
	\label{fig:ISCO_params}
\end{figure}

Figure~\ref{fig:ISCO_params} reveals that all three ISCO parameters grow
monotonically with $b$. The growth of $L_{ISCO}$ and $E_{ISCO}$ becomes more
pronounced at larger $P$, whereas $r_{ISCO}$ is generically reduced as $P$
increases. In the limit $b=0$, the VBH reduces to the SAdS black hole. Relative
to their SAdS values, $b$ systematically enlarges $r_{ISCO}$, $L_{ISCO}$ and
$E_{ISCO}$; test particles thus circle VBHs at larger radii than around SAdS
black holes. This points to a considerable dynamical difference between the two
geometries.

\begin{figure}[tbp]
	\centering
	\includegraphics[width=0.8\textwidth]{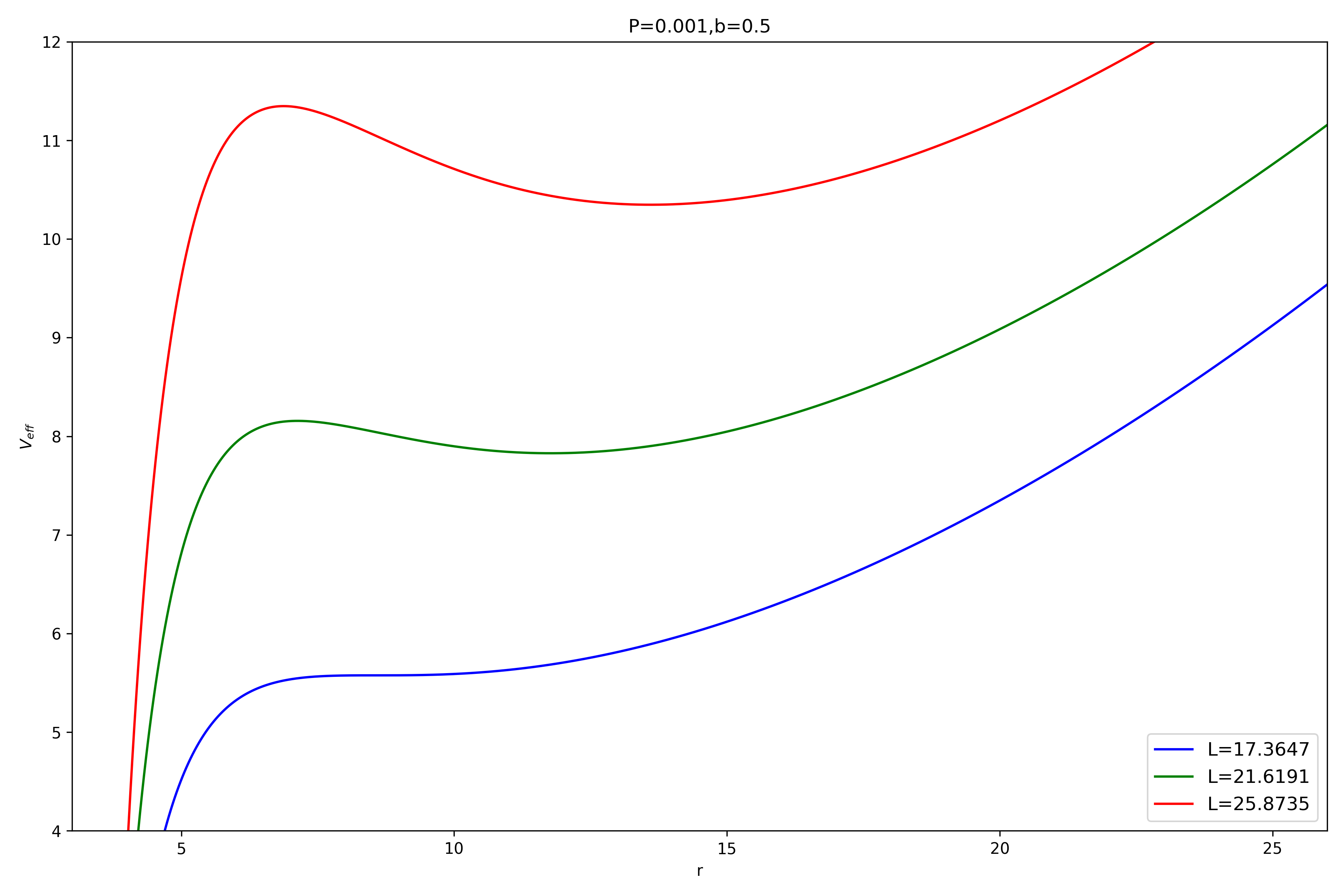}
	\caption{Effective potential vs. $r$ for different angular momenta (with $P=0.001$, $b=0.5$).}
	\label{fig:Veff_L}
\end{figure}

Outside the ISCO, whether the particle remains bound follows from the shape of
$V_{\mathrm{eff}}$ (\ref{eq:effective_potential}), together with the radial
equation of motion (\ref{eq:radial_equation}). As shown in
figure~\ref{fig:Veff_L}, the blue line corresponds to $L = L_{ISCO}$ satisfying
conditions (\ref{eq:ISCO_conditions}). For $L>L_{ISCO}$ the effective potential
develops two extrema, whose separation grows with angular momentum.

\begin{figure}[tbp]
	\centering
	\includegraphics[width=0.8\textwidth]{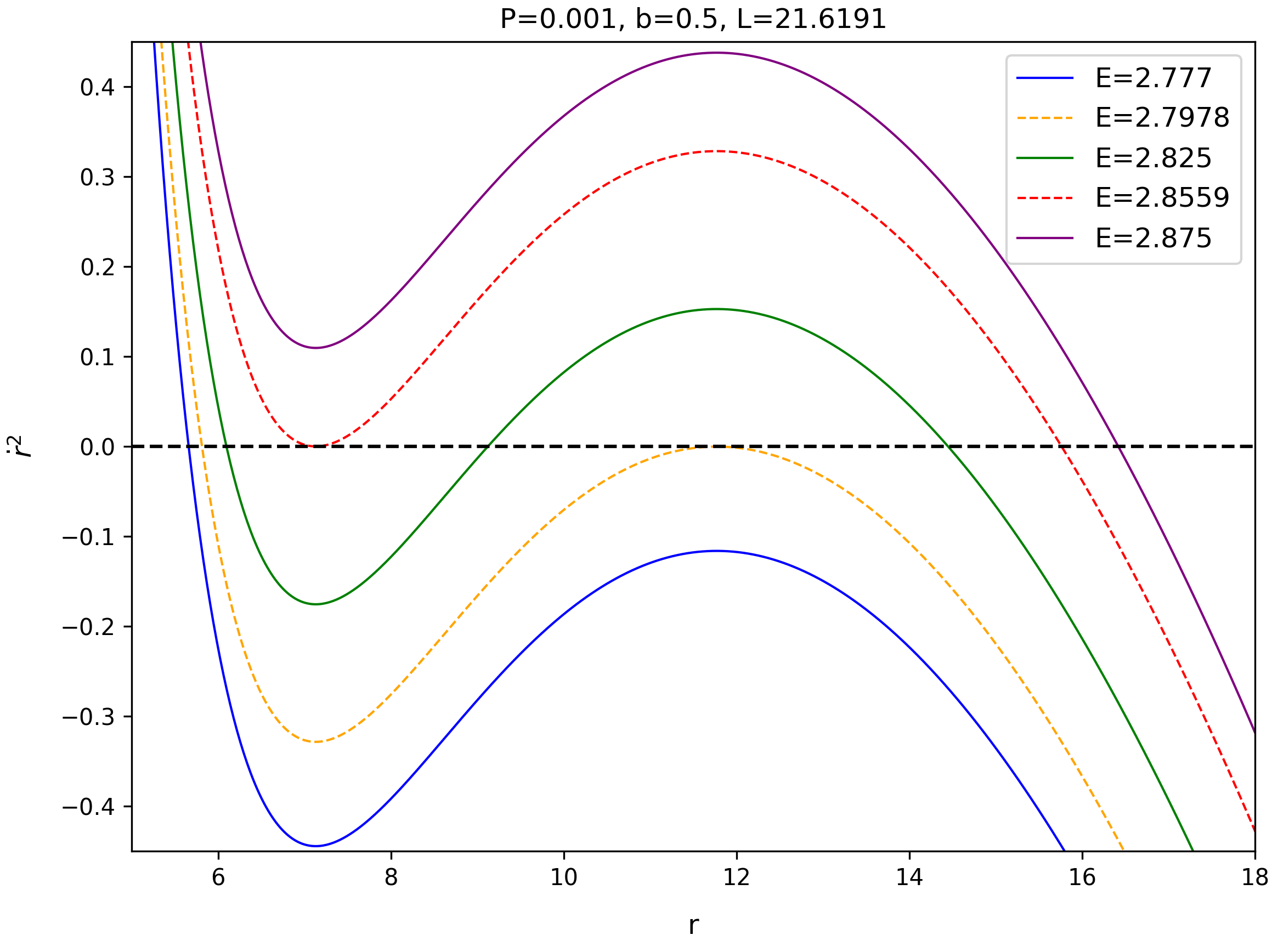}
	\caption{Equation of motion vs. $r$ for different energies (with $P=0.001$, $b=0.5$, and angular momentum $L = 21.6191$).}
	\label{fig:Edot_r}
\end{figure}

In figure~\ref{fig:Edot_r}, for fixed other parameters ($L > L_{ISCO}$), we
select different energies to discuss the test particle's equation of motion and
analyse its bound states. Raising the energy changes the multiplicity of roots of
the radial equation in the sequence: $1\rightarrow2\rightarrow3\rightarrow2\rightarrow1$.
We find that when $E < 2.7978$ (one root), the test particle cannot form a
stable bound orbit and always falls into the black hole; meaningful analysis of
orbital motion is therefore not possible in this regime. When $E = 2.7978$ (two
roots), the expression for ${\dot{r}}^{2}$ is negative between the two roots, so
the particle cannot exist in that region. This energy marks a critical threshold.
In the window $2.7978 < E < 2.8559$ three roots appear, and ${\dot{r}}^{2}$ stays
positive between the two outer ones; bound motion is therefore allowed there.
When $E = 2.8559$ (two roots), similar to the bound-orbit case, the expression
for ${\dot{r}}^{2}$ is positive between the two roots; this energy is likewise
critical. For $E>2.8559$ (one root), as in the first case, no bound orbit is
possible and the particle falls in.

Even though a test particle in AdS can only be bound or captured, $L$ appears
able to grow indefinitely beyond $L_{ISCO}$. However, we find that when the
angular momentum becomes too large, the difference between the two extremum
values of the effective potential $V_{eff}$ exceeds 1 (as shown in
figure~\ref{fig:Veff_L} for $L = 25.8735$). Consequently, once $L$ is large
enough, an energy that would naively admit bound orbits can produce
${\dot{r}}^{2} > 1$ (equivalently ${\dot{r}}^{2} > c^{2}$ in SI units) over part
of the range --- a physically inadmissible outcome. Therefore, we introduce a
critical angular momentum $L_{s}$, corresponding to the case where the difference
between the two extremum values of $V_{eff}$ equals 1. The range $L < L_{s}$ is
considered safe. Although narrowing the $E$ window could in principle make any
$L$ admissible, this brings additional restrictions that we analyse in the next
section. When the angular momentum satisfies $L_{ISCO} < L < L_{s}$, the
corresponding orbits are bound orbits. Since orbits with $L$ too close to
$L_{ISCO}$ are nearly circular and produce less distinct gravitational wave
signals, we subsequently use $L_{m} = (L_{ISCO} + L_{s})/2$ for discussion. This
choice exhibits clear features while ensuring safety.
Table~\ref{tab:orbital_parameters} presents $r_{ISCO}$, $E_{ISCO}$, $L_{ISCO}$,
$L_{s}$, and $L_{m}$ for the various parameter combinations, and these values
will be used in the subsequent discussion.

\begin{table}[tbp]
	\centering
	\setlength{\tabcolsep}{9pt}
	\caption{$r_{ISCO}$, $E_{ISCO}$, $L_{ISCO}$, $L_{s}$, and $L_{m}$ for different parameter combinations.}
	\begin{tabular}{|S[table-format=1.4]|S[table-format=1.1]|S[table-format=1.4]|S[table-format=1.4]|S[table-format=1.4]|S[table-format=1.4]|S[table-format=1.4]|}
		\hline
		{$P$} & {$b$} & {$r_{ISCO}$} & {$E_{ISCO}$} & {$L_{ISCO}$} & {$L_{s}$} & {$L_{m}$} \\
		\hline
		0.0001 & 0   & 5.2619 & 0.9809 & 3.7061 & 7.1128 & 5.4095 \\
		& 0.3 & 8.5202 & 1.0378 & 7.7529 & 14.5724 & 11.1627 \\
		& 0.5 & 10.0540 & 1.0873 & 10.1637 & 18.7961 & 14.4799 \\
		& 0.7 & 11.4663 & 1.1458 & 12.6694 & 23.0243 & 17.8469 \\
		& 1.0 & 13.4640 & 1.2485 & 16.6870 & 29.5091 & 23.0981 \\
		\hline
		0.0005 & 0   & 4.6039 & 1.1087 & 4.3147 & 7.9221 & 6.1184 \\
		& 0.3 & 7.5167 & 1.4415 & 10.3026 & 17.3936 & 13.8481 \\
		& 0.5 & 8.9311 & 1.6850 & 14.2066 & 22.9871 & 18.5969 \\
		& 0.7 & 10.2359 & 1.9526 & 18.3468 & 28.6012 & 23.4740 \\
		& 1.0 & 12.0562 & 2.3878 & 24.9159 & 37.0458 & 30.9809 \\
		\hline
		0.001  & 0   & 4.3490 & 1.2543 & 4.8664 & 8.5571 & 6.7118 \\
		& 0.3 & 7.1978 & 1.9075 & 12.4440 & 19.4864 & 15.9652 \\
		& 0.5 & 8.5681 & 2.3613 & 17.3647 & 25.8735 & 21.6191 \\
		& 0.7 & 9.8036 & 2.8361 & 22.4058 & 32.0887 & 27.2473 \\
		& 1.0 & 11.4646 & 3.5559 & 29.9213 & 40.9340 & 35.4277 \\
		\hline
		0.002  & 0   & 4.1407 & 1.5323 & 5.7367 & 9.4904 & 7.6136 \\
		& 0.3 & 6.9359 & 2.7793 & 15.4793 & 22.2678 & 18.8736 \\
		& 0.5 & 8.2212 & 3.5699 & 21.4018 & 29.3023 & 25.3521 \\
		& 0.7 & 9.3246 & 4.3269 & 26.9790 & 35.6431 & 31.3111 \\
		& 1.0 & 10.7126 & 5.3443 & 34.3441 & 43.6962 & 39.0202 \\
		\hline
	\end{tabular}
	\label{tab:orbital_parameters}
\end{table}

\begin{figure}[tbp]
	\centering
	\includegraphics[width=0.8\textwidth]{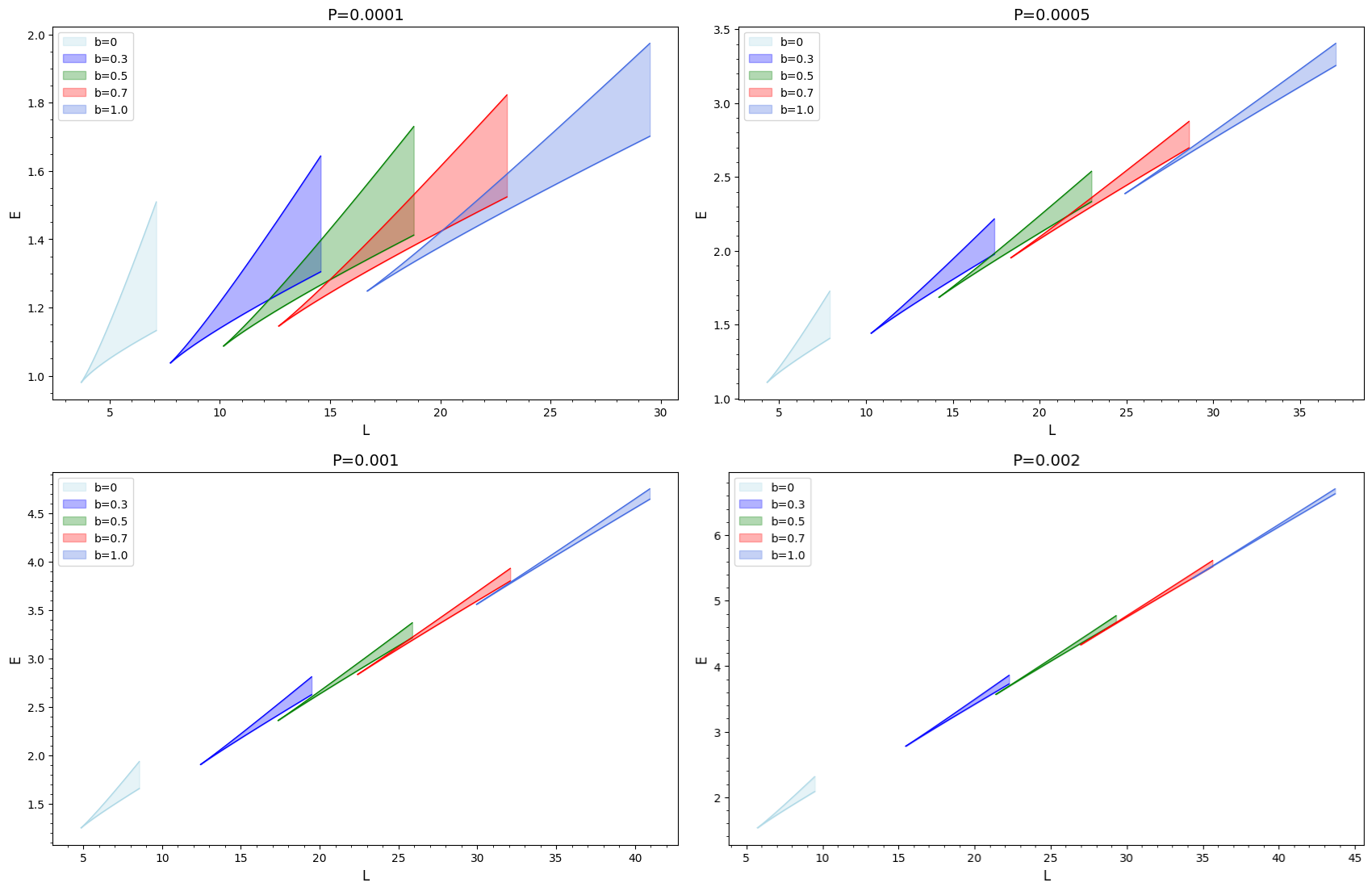}
	\caption{$E-L$ space for different parameter combinations (axis scales differ between plots).}
	\label{fig:EL_space_params}
\end{figure}

\begin{figure}[tbp]
	\centering
	\includegraphics[width=0.8\textwidth]{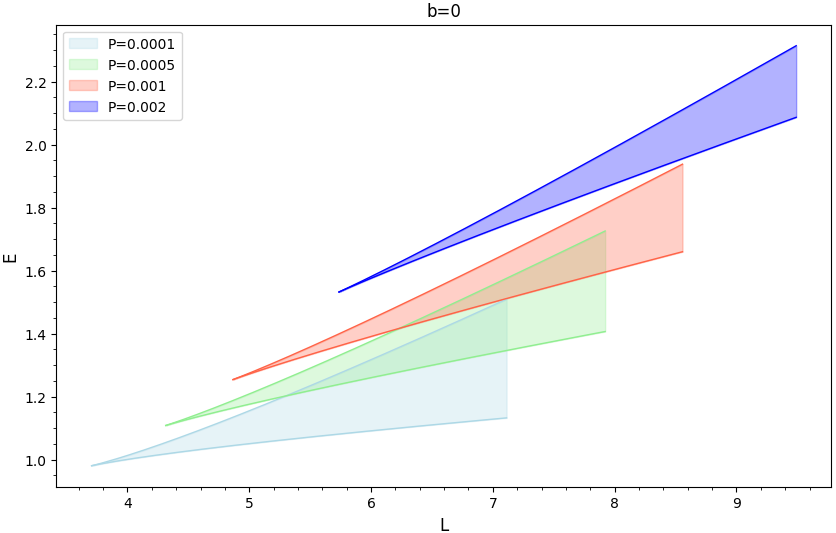}
	\caption{$E-L$ space for different $P$ with $b=0$.}
	\label{fig:EL_space_P}
\end{figure}

Figure~\ref{fig:EL_space_params} shows the $E-L$ space for
$L_{ISCO} < L < L_{s}$ based on the above theory. Although the allowed range of
$E$ for $L_{s}$ decreases with increasing $P$, the prominent features in the
figure mostly reflect the different axis scales (see figure~\ref{fig:EL_space_P}).
In figure~\ref{fig:EL_space_params}, as $P$ gradually increases, the $E-L$ spaces
become elongated and diverge from each other. The difference between the $E-L$
spaces of the VBH and the SAdS black hole is particularly distinct, indicating
that the VBH exhibits more prominent features compared to SAdS black holes.

\section{Influence of VBH Parameters on Periodic Orbit Dynamics}
\label{sec:periodic}

Section~\ref{sec:bound} showed that the VBH values of $r_{ISCO}$, $L_{ISCO}$,
$E_{ISCO}$, and the corresponding $E-L$ region, deviate substantially from the
SAdS case. One may therefore expect the periodic orbits --- and the gravitational
waves they generate --- to carry equally distinctive imprints of these
parameters.

\begin{figure}[tbp]
	\centering
	\includegraphics[width=0.8\textwidth]{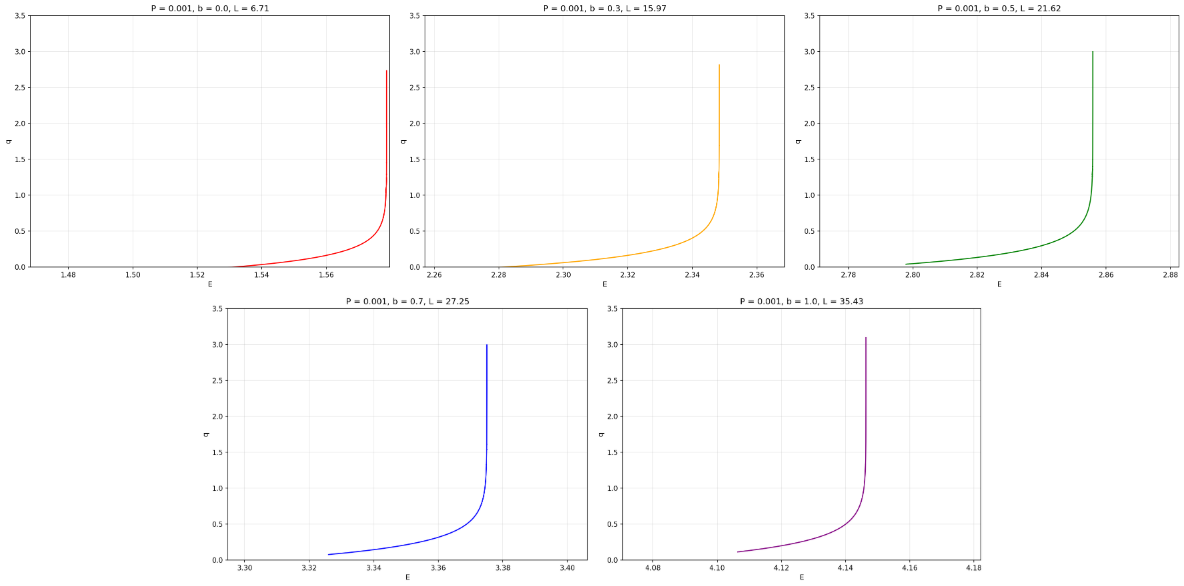}
	\caption{Parameter $q$ diverges as energy increases (with $P=0.001$, angular momentum $L = L_{m}$).}
	\label{fig:q_vs_E}
\end{figure}

Orbits are classified with the scheme of Levin and Perez-Giz \cite{Levin:2008mq},
in which each geodesic carries a number $q$, rational or irrational:
\begin{equation}
	q = \frac{\Delta\phi}{2\pi} - 1 = w + \frac{v}{z}. \label{eq:q_definition}
\end{equation}
Rational $q$ identifies a periodic orbit --- the case of interest here --- whereas
irrational $q$ describes precessing motion, which can itself be approximated by
periodic orbits \cite{Levin:2008mq,Uktamov2026}. The integers $w$, $v$ and $z$
are, respectively, the zoom, vertex and leaf numbers of the orbit in the
terminology of ref.~\cite{Levin:2008mq}. $\Delta\phi$ is the change in angle over
one period, given by:
\begin{equation}
	\Delta\phi = \oint d\phi = 2\int_{r_{1}}^{r_{2}}{\frac{d\phi}{dr}dr}, \label{eq:Deltaphi}
\end{equation}
where $r_{1}$ and $r_{2}$ are the turning points of the test particle motion,
i.e., the latter two roots of ${\dot{r}}^{2} = 0$. Substituting expressions
(\ref{eq:angular_momentum}), (\ref{eq:radial_equation}), and (\ref{eq:Deltaphi})
into (\ref{eq:q_definition}) yields:
\begin{multline}
	q = \frac{\Delta \phi}{2\pi} - 1 = \frac{1}{\pi} \int_{r_1}^{r_2} \frac{d\phi}{dr} \, dr - 1 \\
	= \frac{1}{\pi} \int_{r_1}^{r_2} \frac{L}{r^{2} \sqrt{E^{2} - f(r) \left(1 + \frac{L^{2}}{r^{2}}\right)}} \, dr - 1. \label{eq:q_integral}
\end{multline}
This integral form follows refs.~\cite{Zhao:2024exh,Meng2025}. Since $q$ depends
on both $E$ and $L$, fixing $L = L_{m}$ reduces it to a function of $E$ alone;
figure~\ref{fig:q_vs_E} shows that $q$ grows slowly from 0 and then diverges as
$E$ increases \cite{Levin:2008mq,Uktamov2026}. This means that by choosing
appropriate $E$, we can obtain desired rational $q$. Regarding the method
mentioned earlier of narrowing the range of $E$ to avoid violating the safe
condition when $L$ is too large, we find that using this method requires the
range of $E$ to cover all desired rational $q$. However, due to the integral term
in the calculation of $q$ (eq.~(\ref{eq:q_integral})), this constraint becomes
overly complex. Hence, the artificially introduced $L_{s}$ proves necessary.

Table~\ref{tab:periodic_orbits_energies} lists, for each parameter set and for
$L = L_{m}$, the energy $E$ needed to realise the periodic-orbit types ($z$, $w$,
$v$) of interest, corresponding to desired rational $q$.

\begin{table}[tbp]
	\centering
	\footnotesize
	\setlength{\tabcolsep}{7pt}
	\caption{Energies corresponding to periodic orbits ($z$, $w$, $v$) for different parameter combinations, with $L = L_{m}$.}

	\begin{tabular}{|S[table-format=1.4]|S[table-format=1.1]|S[table-format=1.8]|S[table-format=1.8]|S[table-format=1.8]|S[table-format=1.8]|S[table-format=1.8]|S[table-format=1.8]|}
		\hline
		{$P$} & {$b$} & {$E(1,1,0)$} & {$E(1,2,0)$} & {$E(2,1,1)$} & {$E(2,2,1)$} & {$E(3,1,2)$} & {$E(3,2,2)$} \\
		\hline
		0.0001 & 0   & 1.21978315 & 1.21996390 & 1.21995270 & 1.21996463 & 1.21995984 & 1.21996467 \\
		& 0.3 & 1.31964860 & 1.32000079 & 1.31997498 & 1.32000283 & 1.31999097 & 1.32000293 \\
		& 0.5 & 1.38933196 & 1.38977384 & 1.38973950 & 1.38977674 & 1.38976055 & 1.38977689 \\
		& 0.7 & 1.46627547 & 1.46679516 & 1.46675299 & 1.46679890 & 1.46677864 & 1.46679910 \\
		& 1.0 & 1.59488049 & 1.59549915 & 1.59544643 & 1.59550408 & 1.59547820 & 1.59550436 \\
		\hline
		0.0005 & 0   & 1.39661609 & 1.39679505 & 1.39678403 & 1.39679577 & 1.39679107 & 1.39679581 \\
		& 0.3 & 1.81396443 & 1.81435937 & 1.81432920 & 1.81436187 & 1.81434775 & 1.81436200 \\
		& 0.5 & 2.09882961 & 2.09933538 & 2.09929368 & 2.09933914 & 2.09931896 & 2.09933934 \\
		& 0.7 & 2.40335158 & 2.40394300 & 2.40389154 & 2.40394793 & 2.40392242 & 2.40394821 \\
		& 1.0 & 2.88684570 & 2.88752021 & 2.88745819 & 2.88752653 & 2.88749501 & 2.88752691 \\
		\hline
		0.001  & 0   & 1.57849955 & 1.57869107 & 1.57867905 & 1.57869188 & 1.57868670 & 1.57869191 \\
		& 0.3 & 2.34793762 & 2.34837570 & 2.34834058 & 2.34837877 & 2.34836198 & 2.34837893 \\
		& 0.5 & 2.85534368 & 2.85588797 & 2.85584080 & 2.85589247 & 2.85586913 & 2.85589273 \\
		& 0.7 & 3.37445277 & 3.37506066 & 3.37500527 & 3.37506624 & 3.37503820 & 3.37506657 \\
		& 1.0 & 4.14572581 & 4.14636339 & 4.14630280 & 4.14636979 & 4.14633852 & 4.14637019 \\
		\hline
		0.002  & 0   & 1.90885593 & 1.90907397 & 1.90905967 & 1.90907497 & 1.90906870 & 1.90907501 \\
		& 0.3 & 3.31160564 & 3.31208999 & 3.31204862 & 3.31209387 & 3.31207353 & 3.31209409 \\
		& 0.5 & 4.16346438 & 4.16402105 & 4.16397018 & 4.16402620 & 4.16400041 & 4.16402651 \\
		& 0.7 & 4.96355635 & 4.96412491 & 4.96407119 & 4.96413056 & 4.96410290 & 4.96413091 \\
		& 1.0 & 6.02048636 & 6.02100573 & 6.02095604 & 6.02101102 & 6.02098530 & 6.02101136 \\
		\hline
	\end{tabular}
	\label{tab:periodic_orbits_energies}
\end{table}

Table~\ref{tab:periodic_orbits_energies} indicates that at fixed $L = L_{m}$ the
orbit energy $E$ grows with both $P$ and $b$, and the growth is noticeably
stronger than in the SAdS case. In general, periodic orbits around VBHs demand
higher energies than around SAdS black holes, the gap widening with $P$ or $b$.

\begin{figure}[tbp]
	\centering
	\includegraphics[width=0.8\textwidth]{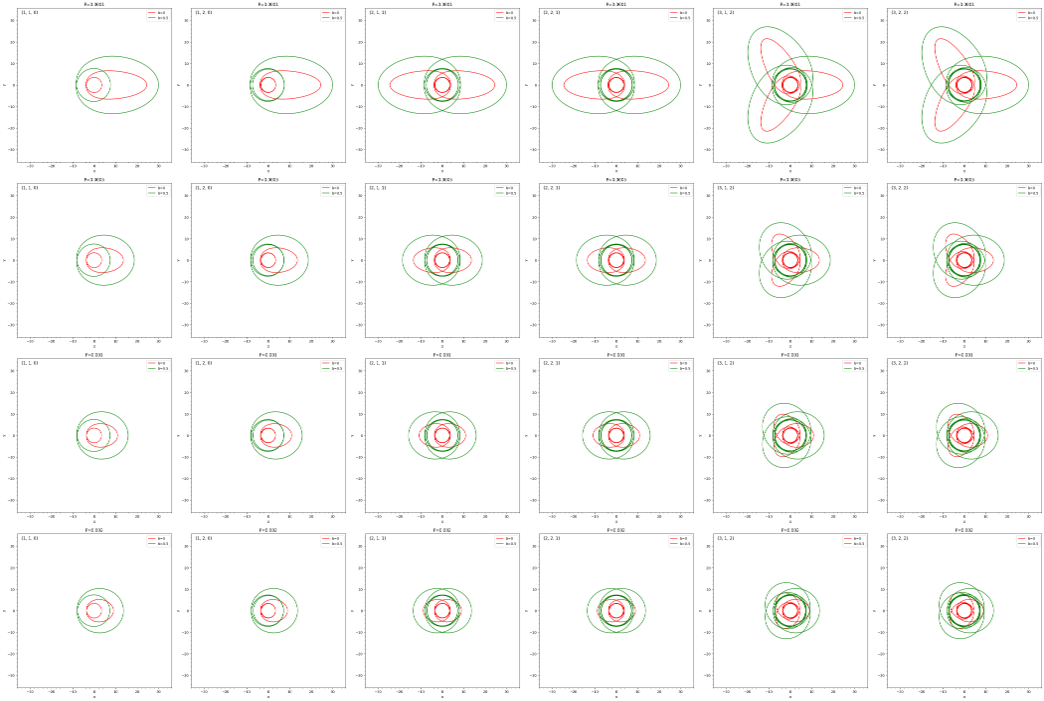}
	\caption{Periodic orbits for SAdS black hole (red) and VBH (green, with fixed $b=0.5$). From top to bottom, $P = 0.0001, 0.0005, 0.001, 0.002$. From left to right, orbit types $(1,1,0)$, $(1,2,0)$, $(2,1,1)$, $(2,2,1)$, $(3,1,2)$, $(3,2,2)$. Using coordinate transformation $(x,y) = (r\cos\phi, r\sin\phi)$.}
	\label{fig:periodic_orbits_comparison}
\end{figure}

\begin{figure}[tbp]
	\centering
	\includegraphics[width=0.8\textwidth]{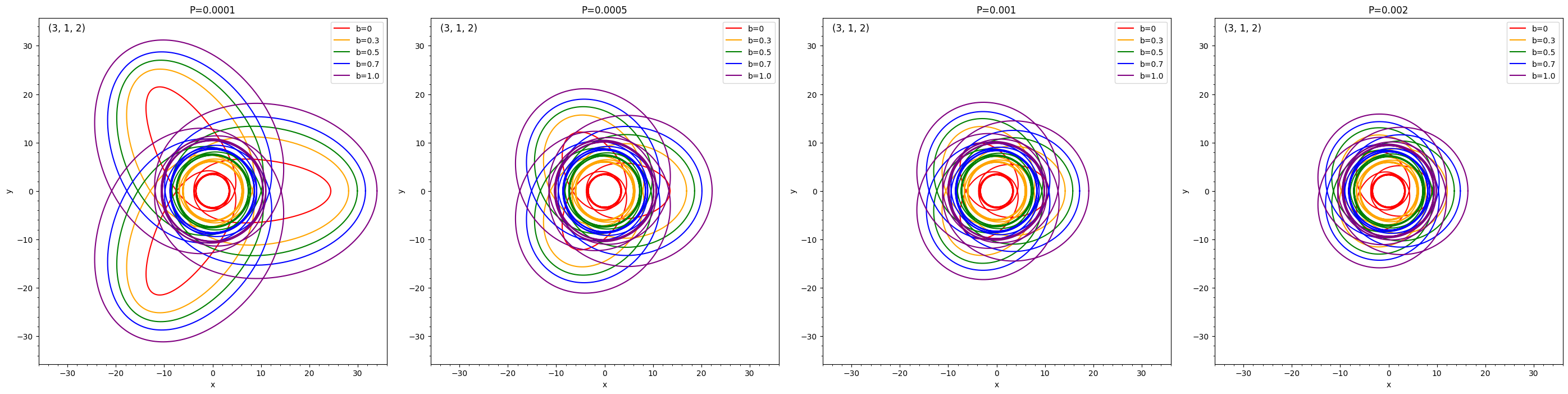}
	\caption{Periodic orbits of type $(3,1,2)$ for SAdS black hole ($b=0$) and the VBH for different $P$. Using coordinate transformation $(x,y) = (r\cos\phi, r\sin\phi)$.}
	\label{fig:periodic_orbits_312}
\end{figure}

Figures~\ref{fig:periodic_orbits_comparison} and~\ref{fig:periodic_orbits_312}
display the ($z,w,v$) periodic orbits of SAdS black holes and VBHs for various
parameter combinations, both with $L=L_m$ and the energies of
table~\ref{tab:periodic_orbits_energies}. Figure~\ref{fig:periodic_orbits_comparison}
confirms the geometric meaning of the labels of \cite{Levin:2008mq}: $w$ counts
zooms, $v$ counts vertices and $z$ counts leaves --- the three integers
characterising the orbit's morphology. Figure~\ref{fig:periodic_orbits_312} shows
that increasing $P$ pulls both the inner and outer branches of the orbit toward
the hole, the outer branch more strongly. For every $P$, the VBH orbit extends
farther from the hole than its SAdS counterpart, and this displacement grows with
$b$.

In short, the thermodynamic parameters $P$ and $b$ also leave a clear imprint on
the periodic-orbit dynamics. This implies that not only can periodic orbit
dynamics be used to distinguish VBHs from SAdS black holes, but they can also be
used to indirectly study related black hole thermodynamics (such as phase
transitions).

\section{Gravitational Wave Radiation from Periodic Orbits in VBH Spacetime}
\label{sec:gw}

For the EMR configuration considered here, the VBH acts as the central
supermassive body and the companion --- modelled as a point-like test particle
since $m\ll M$ --- travels along a periodic orbit deep in the strong-field
region, emitting gravitational waves periodically \cite{Zhao:2024exh}. The
companion's back-reaction on the background geometry is negligible, so over the
few orbital periods of interest the losses of energy and angular momentum remain
small and the adiabatic approximation is justified
\cite{Zhao:2024exh,Glampedakis:2002ya}.

Accordingly, $E$ and $L$ are treated as constants within one period, and only the
radiation emitted over that period is computed --- exactly as done for other EMR
backgrounds in refs.~\cite{Glampedakis:2002ya,Tu:2023xab,Li:2024tld,Zi:2023qfk,Zhao:2024exh,Shabbir:2025kqh},
starting with the zoom-whirl analysis of Glampedakis and Kennefick
\cite{Glampedakis:2002ya}.

Waveforms are generated with the ``Kludge'' prescription of
ref.~\cite{Babak:2006uv}: once the periodic orbit is known
(section~\ref{sec:periodic}), the emitted radiation follows from the quadrupole
formula of refs.~\cite{Maselli:2021men,Liang:2022gdk}, truncated at second order:
\begin{equation}
	h_{ij} = \frac{4\eta M}{D_{L}}\left( v_{i}v_{j} - \frac{m}{r}n_{i}n_{j} \right), \label{eq:quadrupole}
\end{equation}
where $M$ is the mass of the VBH, $m$ is the mass of the test particle, $D_{L}$
is the luminosity distance to the system, $\eta = Mm/(M+m)^{2}$ is the symmetric
mass ratio, $v_{i}$ or $v_{j}$ are components of the velocity vector, and $n_{i}$
or $n_{j}$ are components of the unit radial vector. As in
\cite{Maselli:2021men,Liang:2022gdk}, the wave is projected onto the frame of a
distant observer and written in terms of the two polarisations:
\begin{align}
	&h_{+} = - \frac{2\eta M^{2}}{D_{L}r}\left( 1 + \cos^{2}\iota \right)\cos(2\phi + 2\zeta), \\
	&h_{\times} = - \frac{4\eta M^{2}}{D_{L}r}\cos\iota\sin(2\phi + 2\zeta), \label{eq:polarizations}
\end{align}
here $\iota$ denotes the inclination of the orbital plane with respect to the
observer's line of sight, $\phi$ tracks the orbital phase of the source, and
$\zeta$ is a latitude angle fixing the detector orientation.

To isolate the effect of the VBH parameters, we adopt throughout the benchmark
values of refs.~\cite{Zhao:2024exh,Maselli:2021men,Liang:2022gdk}:
$M=10^{7}M_\odot$ for the VBH, $m=10M_\odot$ for the companion,
$D_L=200\ \mathrm{Mpc}$, $\iota=\pi/4$ and $\zeta=\pi/4$. All numerical work and
plots in this section are carried out in SI units.

\begin{figure}[tbp]
	\centering
	\includegraphics[width=0.8\textwidth]{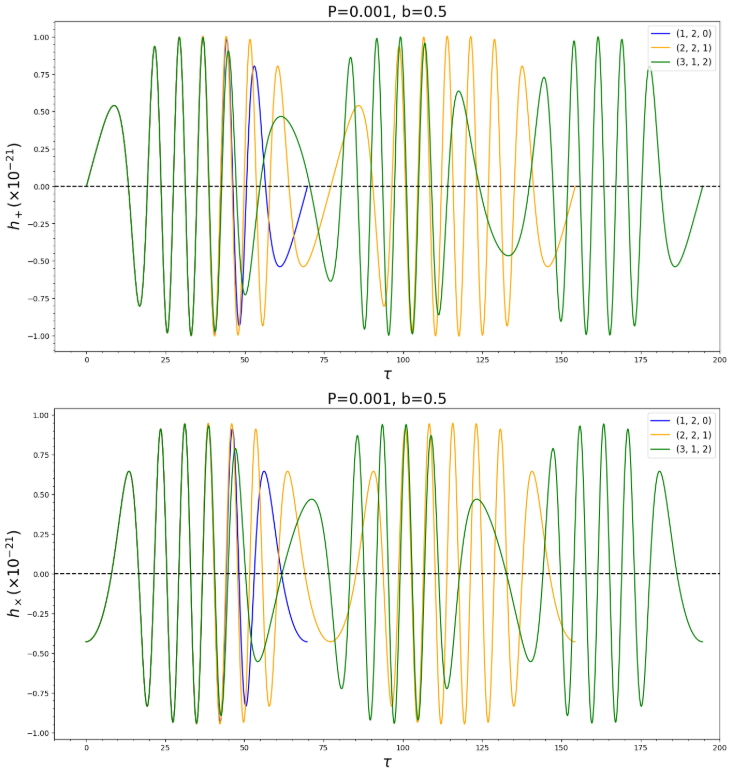}
	\caption{Gravitational waveforms for different periodic orbit types. Upper panel: $h_{+}$ polarization; lower panel: $h_{\times}$ polarization. Fixed parameters $P = 0.001$, $b = 0.5$.}
	\label{fig:waveforms_types}
\end{figure}

\begin{figure}[tbp]
	\centering
	\includegraphics[width=0.8\textwidth]{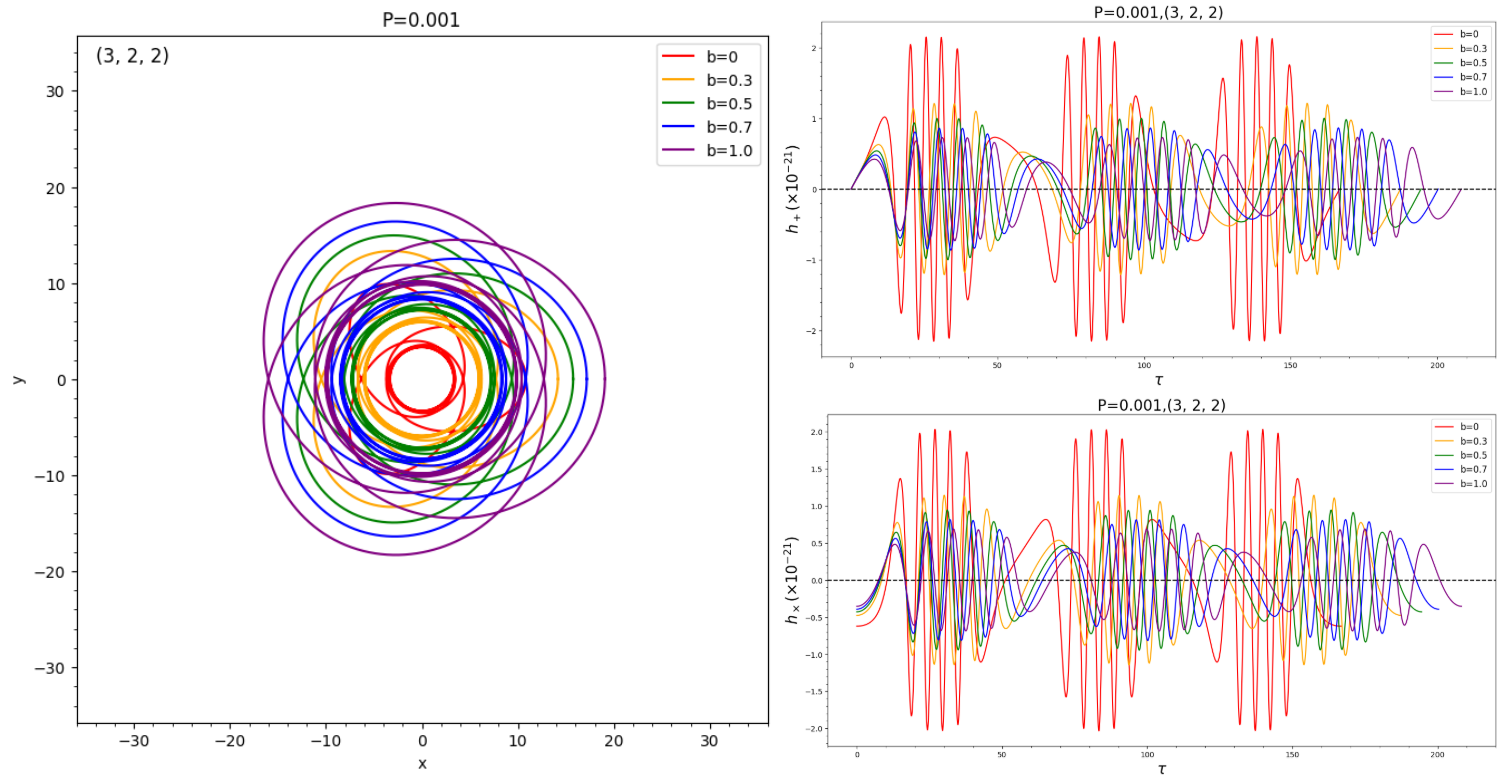}
	\caption{Periodic orbits $(3,2,2)$ and corresponding gravitational waveforms for SAdS black hole ($b=0$) and VBH with different $b$. Fixed $P = 0.001$.}
	\label{fig:waveforms_b_comparison}
\end{figure}

Figure~\ref{fig:waveforms_types} presents the waveforms, for both $h_{+}$ and
$h_{\times}$ polarizations, produced by three representative orbit types. In
every trace a long, slowly oscillating, low-amplitude stretch (the zoom phase,
during which the particle stays far from the centre) is followed by a short,
rapidly oscillating, high-amplitude burst (the whirl phase, executed near the
hole).

The role of $b$ is illustrated in figure~\ref{fig:waveforms_b_comparison}, where
$P$ is fixed and the $(3,2,2)$ orbit together with its waveform are compared
between the SAdS case ($b=0$) and VBHs with varying $b$. Relative to the SAdS
benchmark, VBH waveforms last longer in both the zoom/whirl phases and the full
period, while their oscillation amplitude is reduced. This contrast becomes more
pronounced as $b$ increases. In short, the waveforms depend strongly on $b$: as
$b$ grows the period lengthens while the amplitude is suppressed. This suggests
that future gravitational wave detections could potentially identify VBHs or
distinguish them from SAdS black holes. These wave characteristics therefore
offer a remote probe of the underlying thermodynamic parameters and phase
behaviour.

However, while this method has performed well in studies such as
refs.~\cite{Zhao:2024exh,Shabbir:2025kqh} and suffices to tell SAdS black holes
from VBHs, it still rests on the adiabatic approximation and omits higher-order
multipoles. This leads to some information loss and potential inaccuracies in the
theoretical model. Future research should incorporate more precise waveform
models to study the influence of $b$ on gravitational waveforms more deeply and
to revise the theoretical model.

\section{Conclusion}
\label{sec:conclusion}

This work has examined the VBH family --- asymptotically AdS black holes whose
extended-phase-space thermodynamics reproduces exactly that of a van der Waals
fluid \cite{Rajagopal:2014ewa,Delsate:2014zma}. With suitable $P$ and $b$, all
energy conditions can be satisfied near the horizon. Treating the VBH as the
central supermassive object of an EMR binary and the light companion as a test
particle, we analysed the timelike geodesics and the gravitational-wave
signatures of periodic orbits, focusing on the imprint of $b$. Our main findings
are as follows.

With the critical value $L_s$, the bound orbits lie between the ISCO and the
$L_s$ orbit. The ISCO radius, energy and angular momentum are all sensitive to
$P$ and $b$ (figure~\ref{fig:ISCO_params}): they grow with $b$, and for
$L_{ISCO}$ and $E_{ISCO}$ the growth is amplified at larger $P$. Since $P$ is the
thermodynamic pressure, a larger $P$ generally results in smaller values of
$r_{ISCO}$ overall, without significantly altering the influence of $b$. The
limit $b=0$ restores the SAdS black hole. Overall, the presence of $b$ makes
$r_{ISCO}$, $L_{ISCO}$, and $E_{ISCO}$ larger for VBHs than for SAdS black holes.
This conclusion also holds for critical orbits and bound orbits. Both parameters
likewise reshape the bound-orbit $E-L$ plane (figures~\ref{fig:EL_space_params}
and~\ref{fig:EL_space_P}): a larger $b$ shifts it towards higher $E$ and $L$,
while a larger $P$ stretches and separates the allowed regions, so that at large
$P$ and $b$ the VBH $E-L$ space departs clearly from the SAdS one.

Considering $L_{s}$ and $L_{ISCO}$, we focused on bound orbits corresponding to
$L_{m} = (L_{ISCO}+L_{s})/2$ and used the orbit classification method to encode
periodic orbits. For fixed $L = L_{m}$, $q$ increases slowly then rapidly diverges
as $E$ increases (see figure~\ref{fig:q_vs_E}). Test particles orbiting VBHs need
higher energies than around the SAdS black hole ($b=0$).
Figures~\ref{fig:periodic_orbits_comparison} and~\ref{fig:periodic_orbits_312}
compare periodic orbit images, showing clear differences, especially for larger
$b$, making it possible to distinguish VBHs from SAdS black holes via periodic
orbits. Furthermore, treating the VBH as the central black hole in an EMR system,
applying the adiabatic approximation over a single period, and neglecting
higher-order multipoles yields gravitational waves in both polarizations (see
figures~\ref{fig:waveforms_types} and~\ref{fig:waveforms_b_comparison}). Zoom and
whirl signatures are already distinguishable within a single period of the
waveform. For fixed $P$, a larger $b$ results in longer periods and smaller
amplitudes. Overall, this work attempts to link black hole thermodynamic
parameters with the dynamical behavior of corresponding EMR systems, intuitively
and quantitatively demonstrating the influence of the thermodynamic pressure $P$
and the Van der Waals parameter $b$. This provides an approach for studying black
hole thermodynamics through dynamical features such as periodic orbits and
gravitational wave signals.

In future research, we will attempt to refine the model, for example, by
introducing spin or charge to satisfy stricter energy conditions. Moreover, due
to their immense mass and tiny scale, black holes involve both general relativity
and quantum theory. A further caveat is that the construction relies on the van
der Waals-type criticality shared by these AdS black holes and real fluids;
generalisations built on Bose-Einstein or Fermi-Dirac statistics may ultimately
be necessary. Extensions along the lines of refs.~\cite{Anand:2025ttx,Oubagha:2023vcc}
could incorporate the generalised/extended uncertainty principle or rainbow
gravity and probe how such quantum-gravity corrections modify the orbital
dynamics. More accurate waveform models will likewise be pursued in future work.

%


\end{document}